# Reversible Data Hiding in Encrypted Images by Lossless Pixel Conversion

Zi-Long Liu, Chi-Man Pun

*Abstract*—**Reversible data hiding in encrypted image (RDHEI) becomes a hot topic, and a lot of algorithms have been proposed to optimize this technology. However, these algorithms cannot achieve strong embedding capacity. Thus, in this paper, we propose an advanced RDHEI scheme based on lossless pixel conversion (LPC). Different from the previous RDHEI algorithms, LPC is inspired by the planar map coloring question, and it performs a dynamic image division process to divide the original image into irregular regions instead of regular blocks as in the previous RDHEI algorithms. In the process of LPC, pixel conversion is performed by region; that is, pixels in the same regions are converted to the same conversion values, which will occupy a smaller size, and then the available room can be reserved to accommodate additional data. LPC is a reversible process, so the original image can be losslessly recovered on the receiver side. Experimental results show that the embedding capacity of the proposed scheme outperforms the existing RDHEI algorithms.**

## I. INTRODUCTION

Reversible data hiding (RDH) requests that the recovered cover is lossless compared with the original cover. It is usually used for tampering certification and non-distortion recovery of multimedia data, especially for scenarios that have high-performance requirements of confidentiality and fidelity, such as law forensics, military imagery, and medical imagery.

A number of algorithms for RDH have been proposed in recent years. In [1], Fridrich *et al.* losslessly compressed image bit planes; thus, spare space can be saved for embedding additional data. Fridrich *et al.* also designed image formats (such as BMP, JPEG, GIF, and PNG) based general framework of RDH in [2]. In [3], Celik *et al.* modified the generalization least significant bit to embed additional data. Tian, in [4], proposed an RDH algorithm based on difference expansion, in which the correlation of adjacent pixels in the natural image is extended. In [5], an RDH algorithm based on histogram shifting is proposed, in which spare space is saved for additional data by shifting the bins of a grayscale histogram of an image. According to the principles of using, the RDH schemes related above can be classified into three categories: lossless compression appending based schemes [1]-[3], difference expansion based schemes [4] and histogram shifting based schemes [5]. There are also some RDH methods extended or combined the three categories to reach better performance, such as [6]-[12].

By transforming the original cover into unrecognized ciphertext, encryption [13] provides confidentiality for image content. However, in some situations, only encryption is not enough. Considering the scenario of a medical image data center, a doctor encrypts a medical image to protect the privacy of his patient. A server of the data center does not know the image content, but she/he wants to manage this encrypted medical image by embedding some notations into it. As the medical image is important for the patient, the embedded encrypted image should be able to be recovered losslessly. In such a case, reversible data hiding in encrypted image (RDHEI) is desired. The combination of encryption and RDH can realize privacy protection, copyright protection, content integrity authentication and ciphertext management of multimedia data.

Some attempts about RDHEI have been made. In [14], Puech *et al.* embedded each additional bit into a block of pixels encrypted by the AES (Advanced Encryption Standard). However, the PSNR of the decrypted image containing additional data declines significantly as the disturbance of additional data. In [15], X. Zhang encrypted the original image by an exclusive-or operation with pseudo-random bits, and then divided the encrypted image into several blocks. The additional data are embedded by flipping three LSBs of the half of pixels in each block. With the help of spatial correlation in natural images, the embedded data can be extracted and the original content can be retrieved. By using syndromes of a parity-check matrix, X. Zhang [16] compressed the LSBs of the encrypted image to vacate room for additional data. X. Zhang also made data extraction independent of image decryption in [16], and the receiver side still used the spatial correlation of decrypted images to recover the original image. Hong *et al.* [17] improved X. Zhang's algorithm of [15] by further exploiting the spatial correlation with a better estimation equation and side match technique at the decoder side to achieve better performance. In [18] and [19], the authors proposed the idea of reserving room before encryption (RRBE). They used a traditional RDH algorithm to reserve room before the image was encrypted, which made the data embedding much easier and could also enhance embedding rate. In [20], the original content was firstly encrypted, then the data hider used LDPC code to compress a half of the fourth LSB in the encrypted image. Finally, the half of the fourth LSB could be used to accommodate the compressed data and the additional data. Cao

Zi-Long LIU and Chi-Man PUN are with the Department of Computer and Information Science, Faculty of Science and Technology, University of Macau, Taipa, Macau S.A.R. 999078, China.

*et al.* [21] used a patch-level sparse representation technique to generate a larger vacated room, which could be used to accommodate additional data. In [22], Qin *et al.* used an analogous stream-cipher and block permutation method to encrypt blocks of the original image, then the encrypted blocks can be classified into smooth blocks and complex blocks. Finally, additional data can be embedded by compressing LSBs of the smooth blocks. Li *et al.* [23] used a block permutation, and a stream cipher combined encryption method and bit replacement in prediction error to improve the embedding rate. Bit plane partition method was proposed, in [24], to realize high capacity RDHEI algorithm. In [25], Liu *et al.* proposed to realize RDHEI by transferring redundant space from the original image to the encrypted image. Furthermore, Liu *et al.*, in [26], proposed another scheme of RDHEI by reversible image reconstruction, which means rearranging the original image to construct a redundancy image. Moreover, new steganography algorithms such as [27]-[30] were also proposed. However, the embedding capacity of the above algorithms still needs to be improved.

In this paper, we propose an advanced RDHEI scheme by lossless pixel conversion (LPC). First of all, LPC is employed to reserve room in the original image. In LPC, redundancy in the original image is first collected into divided blocks. Then, for each divided block, we split it into regions, find out the interconnection relationship among its all regions, and set conversion value for the pixels of its all regions to ensure that the pixels in adjacent regions have different conversion values. The conversion value of a pixel is smaller than its original value, so the available room can be vacated to accommodate additional data. In the second step of the proposed scheme, we assemble several kinds of data together to form the reserved image. Third, we encrypt the reserved image with the traditional standard stream cipher encryption method. Next, with the available room in the encrypted image, additional data can be embedded. Finally, depending on the permissions an authorized receiver has, she/he can either extract additional data or recover image, or both.

The proposed scheme is similar to the previous algorithms, such as [28]-[30]. However, the proposed scheme also contains distinctive contributions, and this can be reflected in the following three points.

- The proposed scheme is inspired by the planar map coloring question, and it generates available room by using four conversion/color values to mark the original image. This idea is novel and has not been proposed by previous algorithms.
- Different from the previous algorithms, the proposed scheme finally divides the original image into irregular regions instead of regular blocks. This can make full use of the spatial correlation of natural images and further improve the performance.
- Image division in the proposed scheme is a dynamic process, which has also not appeared in the previous algorithms. Dynamic process means: according to region merging, after image division, small regions may still be merged into large regions. This dynamic process can optimize the image division results.

The rest of this paper will be organized in the following way. Section II introduces the principle of LPC, and the proposed scheme will be elaborated in section III. In section IV, we will display and analyze the experimental results of the proposed scheme. Finally, the conclusion portion is given in section V.

## II. THE PRINCIPLE OF LOSSLESS PIXEL CONVERSION

In [31]-[32], the planar map coloring question has been proven, and its content is that any planar map can be colored with at most four colors in such a way that no two adjacent regions have the same color. Here, adjacent regions refer to two regions with a common point that is not a corner, which is a kind of point that if and only if it belongs to three or more regions.

Inspired by the planar map coloring question, the principle of LPC is as follows. First, we see an image as a "planar map", the adjacent pixels that have the same value make up regions of this "planar map". Second, we try to mark this "planar map" with at most four conversion values and make sure the adjacent regions of this "planar map" have different conversion values. Third, the marking operation needs two bits to represent the four conversion values, so if the image has more than two bit planes, theoretically, the available room can be reserved by LPC. On the other hand, LPC is reversible; that is, the original pixel values can be losslessly restored. The conversion values identify the location of each region, and all the pixels in the same region have the same original value, so by recording the unique original pixel value of each region, the original image can be recovered without loss.

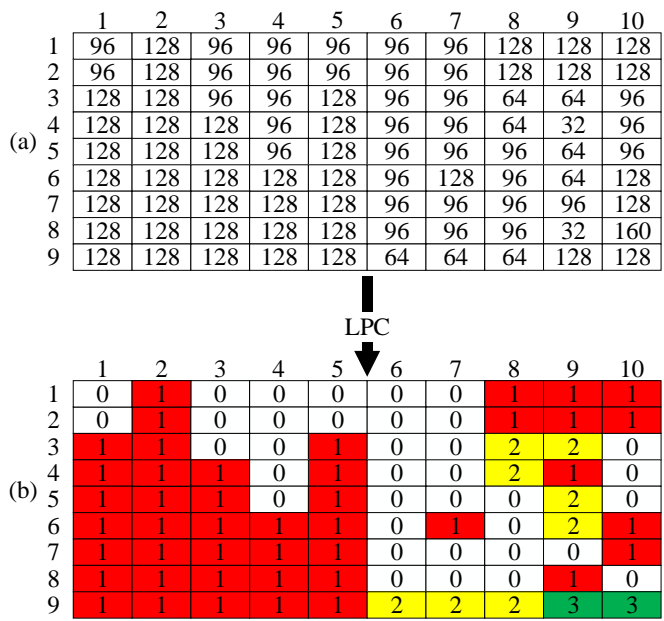

Fig. 1. Example of LPC. Adjacent regions are distinguished by different colors.

For example, Fig. 1(a) is redundant values of some pixels extracted from an arbitrary grayscale image, in which abscissa is labeled on the top and ordinate is labeled on the left. In the process of LPC, we first consider Fig. 1(a) as a "planar map", and adjacent pixels with the same value (such as pixels (8, 3), (8, 4) and (9, 3)) will combine into regions of this "planar map". Second, we use four different conversion values (0, 1, 2, and 3) to mark all the pixels, and the conversion result is shown in Fig. 1(b). As can be seen, pixels in adjacent regions have different conversion values. Third, it is obvious that the conversion value is much smaller than the original value and takes up less room, so by storing the conversion value of each pixel instead of its original value, the available room can be freed up. As a demonstration, to recover Fig. 1(a) from Fig. 1(b), we first divide Fig. 1(b) into regions, then Fig. 1(a) can be recovered by using the recorded unique original pixel values to replace the conversion values in each region.

## III. THE PROPOSED SCHEME

Reversible data hiding in encrypted image (RDHEI) is usually used for tampering certification and non-distortion recovery of multimedia data, especially for scenarios that have high-performance requirements of confidentiality, fidelity, and algorithms such as[33-40]. According to the foregoing analysis, we find that the previous RDHEI algorithms cannot achieve high embedding capacity. So, in this section, we propose our scheme, which will have a better performance. The framework of the proposed scheme is shown in Fig. 2. As we can see, the original image is first processed by LPC (the first four phases) to generate available room, then the reserved image is formed by assembling four kinds of data together. After the reserved image is encrypted, the data owner can embed additional data into the encrypted image, and then the marked encrypted image is generated. Finally, data extraction and image recovery on the receiver side are separable operations.

### A. Lossless pixel conversion

In the proposed scheme, the original image is first processed by LPC, which contains four phases: image pre-processing, region division, region connection detection, and pixel conversion. The details of these four phases are as follows.

#### 1) Image pre-processing

The motivation of this phase contains two points. First, increase the security of the proposed scheme. By this phase, the information contained in the original image will become invisible, so privacy in the original image will be better protected. The second point is to better exploit redundant space. Through this phase, redundant space contained in the original image will be transferred to divided blocks, so this redundancy can be more easily utilized.

In this phase, the bit order of each pixel in the original image will be inversed first, and then the inversed image will be divided. This can be realized by the following three steps.

First, assuming there is an arbitrary pixel $I(a, o)$ in an 8-bit grayscale image $I$ with grayscale value $X$, we do the inversion operation to it by

$$I'(a, o) = \sum_{k=0}^{7} \left( \left\lfloor \frac{X}{2^k} \right\rfloor \bmod 2 \right) \times 2^{7-k} , \tag{1}$$

where $I'(a, o)$ is the inversed pixel of $I(a, o)$. After all the pixels in $I$ is processed by the above step, we can get the inversed image $I'$, the content of which is chaotic. In order to easily utilize the redundancy contained in $I'$, we continue to do the following two steps.

Second, extract $I'_l(a, o)$ that contain the $\lambda$ low order bits of $I'(a, o)$ by

$$I'_l(a, o) = \sum_{k=0}^{\lambda-1} I'(a, o)(k) \times 2^k . \tag{2}$$

The remaining value of $I'(a, o)$ can be calculated by

$$I'_h(a, o) = I'(a, o) - I'_l(a, o) . \tag{3}$$

By using Eq. (2) and Eq. (3) to process all the pixels of $I'$, we can get the $\lambda$ low order bits image $I'_l$, which contains almost all redundancy of $I$, and $8-\lambda$ high order bits image $I'_h$.

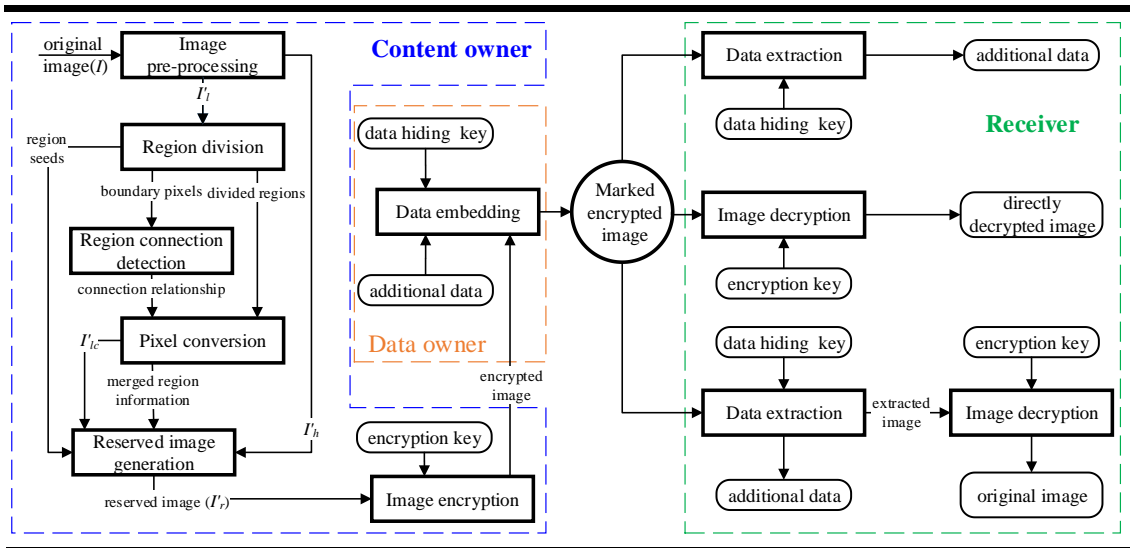

Fig. 2. The framework of the proposed scheme.

Third, divide $I'_l$ into blocks. Assuming the size of $I$ is $M\times N$, and the size of the divided block is $\tau\times\tau$, then arbitrary divided block $B\_I'_l(i, j)$ ($1 \leq i \geq M/\tau$, $1 \leq j \geq N/\tau$) of $I'_l$ can be given by

$$B\_I'_i(i, j) = I'_i((i-1) \times \tau + 1 : i \times \tau, (j-1) \times \tau + 1 : j \times \tau). \quad (4)$$

In $B\_I'_i(i, j)$, an arbitrary pixel whose position coordinate is $(a, o)$ has 4 adjacent pixels, and their coordinates are $(a, o+1)$, $(a, o-1)$, $(a+1, o)$, and $(a-1, o)$, which we name as surrounding pixels of this arbitrary pixel. The following three phases will sequentially process $B\_I'_i(i, j)$ step by step.

### 2) Region division

In order to make fully use of the redundancy in $B\_I'_i(i, j)$, this phase divides it by its spatial correlation. This is to say, adjacent pixels that have the same grayscale value will be divided into a region. Here, this same grayscale value is called the seed of this region. To divide $B\_I'_i(i, j)$, a matrix is needed to label which pixel is processed and which is not. We name this matrix as label matrix. If a pixel is added to a region, the same position element in the label matrix will be set to 1; otherwise, its value will be 0.

We divide $B\_I'_i(i, j)$ into regions according to the following five steps.

**Step 1:** find a seed. This step scans label matrix in raster order that is from left to right and from top to bottom and finds the first element whose value is 0, then the corresponding pixel in $B\_I'_i(i, j)$ will be the seed (represented by $S$) that we are looking for.

**Step 2:** add $S$ to a new region (represented by $R_0$).

**Step 3:** scan the surrounding pixels of $S$, and add the scanned pixels that have the same grayscale value with $S$ to $R_0$.

**Step 4:** repeat doing step 3 to all the pixels that have been added to $R_0$ until there is not new pixel can be added to $R_0$, then region $R_0$ is generated.

**Step 5:** repeat doing the above four steps until all the elements in the label matrix have the value of 1, then $B\_I'_i(i, j)$ is divided.

For example, if the pixels in Fig. 1(a) make up a divided block, then Fig. 3 shows its region division result. As we can see, Fig. 1(a) is divided into 14 regions ($r_0$-$r_{13}$), and seeds for region $r_0$-$r_{13}$ respectively are 96, 128, 96, 128, 64, 96, 32, 64, 128, 128, 32, 160, 64, 128. When a region is generated, the boundary pixels and seed of it should be recorded at the same time. Boundary pixels are the pixels adjacent to the neighboring regions.

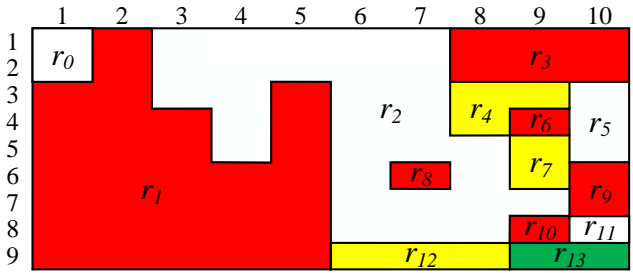

Fig. 3. Region division result of Fig. 1(a).

### 3) Region connection detection

After $B\_I'_i(i, j)$ is divided, regions are generated. In order to mark these regions with four conversion values and make sure the adjacent regions have different conversion values, we should know how a region is connected to other regions. So, in this phase, we use boundary pixels to detect the connection relationship among different regions of $B\_I'_i(i, j)$. Boundary pixel has a feature that boundary pixel itself belongs to the current region and certain surrounding pixels of this boundary pixel belong to adjacent regions of the current region.

Assuming there is an arbitrary region $R_1$ in $B\_I'_i(i, j)$, and $BP$ are boundary pixels of $R_1$, the way we find adjacent regions of $R_1$ is as follows.

**Step 1:** for an arbitrary region (represented by $R_2$) of $B\_I'_i(i, j)$, compare the surrounding pixels of $BP$ with $R_2$. If the two have an intersection, then this $R_2$ is an adjacent region of $R_1$.

**Step 2:** repeat doing step 1 until all the regions in $B\_I'_i(i, j)$ have been compared with the surrounding pixels of $BP$. Then all the adjacent regions of $R_1$ can be found.

For each region in $B\_I'_i(i, j)$, do the above two steps, then an adjacent region list of $B\_I'_i(i, j)$ will generate in this phase. For example, for the regions in Fig. 3, their region connection detection result is shown in Table I.

TABLE I ADJACENT REGION LIST OF THE REGIONS IN FIG. 3.

| Region | Adjacent regions |
|---|---|
| $r_0$ | $r_1$ |
| $r_1$ | $r_0$, $r_2$, $r_{12}$ |
| $r_2$ | $r_1$, $r_3$, $r_4$, $r_7$, $r_8$, $r_9$, $r_{10}$, $r_{12}$ |
| $r_3$ | $r_2$, $r_4$, $r_5$ |
| $r_4$ | $r_2$, $r_3$, $r_5$, $r_6$ |
| $r_5$ | $r_3$, $r_4$, $r_6$, $r_7$, $r_9$ |
| $r_6$ | $r_4$, $r_5$, $r_7$ |
| $r_7$ | $r_2$, $r_5$, $r_6$, $r_9$ |
| $r_8$ | $r_2$ |
| $r_9$ | $r_2$, $r_5$, $r_7$, $r_{11}$ |
| $r_{10}$ | $r_2$, $r_{11}$, $r_{13}$ |
| $r_{11}$ | $r_9$, $r_{10}$, $r_{13}$ |
| $r_{12}$ | $r_1$, $r_2$, $r_{13}$ |
| $r_{13}$ | $r_{10}$, $r_{11}$, $r_{12}$ |

### 4) Pixel conversion

In order to reserve available room in $B\_I'_i(i, j)$, this phase marks all the pixels in it with four conversion values (0, 1, 2, and 3), and ensures pixels in adjacent regions have different conversion values. If pixels in two or more adjacent regions have the same conversion value, we name this situation as conversion collision. In the beginning, the conversion values of all the pixels in $B\_I'_i(i, j)$ is set to a different value, for example, -1. Suppose the initial value of $x$ is 0, an arbitrary region (represented by $R_3$) can be converted through the following four steps.

**Step 1:** assume converting $R_3$ with conversion value $x$.

**Step 2:** check adjacent regions of $R_3$. If conversion collision happens, let $x = x + 1$ and jump to step 1, else jump to step 3.

**Step 3:** if $x$ is less than or equal to 3, then $R_3$ can be converted with conversion value $x$. Otherwise, the case will be that $R_3$ cannot be converted with four conversion values. In this case, we backtrack and modify the conversion value of the recently converted regions, set $x$ to 0, jump to step 1, and try again. This backtracking method can solve most of the conversion collisions. For the conversion collisions that cannot be solved by the backtracking method, region merging method, which will be described below, will be an effective solution. By backtracking and region merging, $R_3$ can be converted, and no adjacent regions have the same conversion value.

**Step 4:** for each pixel in $R_3$, store the converted value of $R_3$ in its two high order bits and set the remaining bits to zero.

In some images, complex textures will cause a small number of conversion collisions that cannot be solved by the above backtracking method. Fortunately, we design the method of region merging as follows to solve this case. Region merging is actually a micro adjustment to the results of region division. Assuming the two participants of region merging are $B'$ (be merged region) and $B''$ (merge region). The main idea of region merging method contains three points.

- The principles of selecting the two merging regions are that the size of $B'$ should be as small as possible and the best choice of $B''$ is preferably a converted region with a larger size. Here, region size means how many pixels it contains. These principles can reduce the solution spending to a minimum in most cases.
- During the region merging, the values of all the pixels in $B'$ will be replaced by the seed value of $B''$.
- Pixels value and position coordinate, which is named merged region information, of all the pixels in $B'$ will be recorded for image recovery.

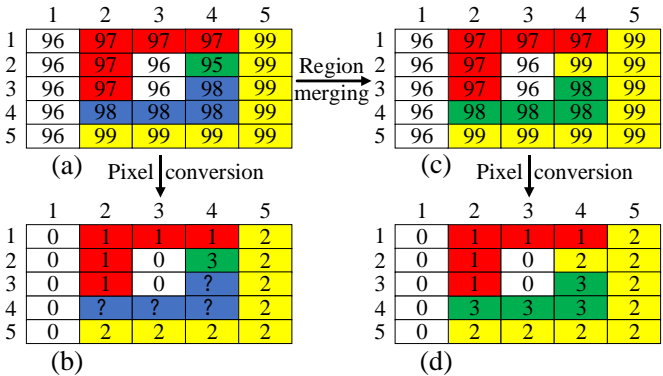

Fig. 4. Example of region merging. Adjacent regions are distinguished by different colors.

An example of using region merging is illustrated in Fig. 4. Suppose Fig. 4(a) is a divided block and the top and left are abscissa and ordinate respectively. As we can see from Fig. 4(b), if we operate pixel conversion directly on Fig. 4(a), the region in blue cannot be converted. Fortunately, region merging will work in this case. According to the first point of region merging idea, the green region in Fig. 4(a) will be selected as $B'$ (be merged region) and the yellow region in Fig. 4(a) will be selected as $B''$ (merge region). Then, the value of the only pixel in the green region is replaced by the seed value of the yellow region. Moreover, pixels value and position coordinate, which are respectively 95 and (4, 2), of the only pixel in the green region will be recorded. Finally, Fig. 4(c) is the divided block processed by region merging. By operating pixel conversion to Fig. 4(c), we obtain Fig. 4(d), in which all the regions can be converted.

After this phase, all the regions in $B\_I'_l(i, j)$ can be converted, and available room can be reserved in the low order bit planes of $B\_I'_l(i, j)$. After all the divided blocks in $I'_l$ are processed by this phase, the converted $\lambda$ low order bit planes $I'_{lc}$ is generated.

### B. Reserved image generation

After the pixel conversion, the reserved available room is distributed in each converted divided block, so we perform this step to obtain a continuous available room. We first encode some auxiliary information into recovery information, and then we produce the reserved image, which contains continuous available room. Finally, we store some control parameters into the reserved image to facilitate data embedding, data extraction, and image recovery.

First, for image recovery, we design two kinds of auxiliary information: unused divided block information and used divided block information, which will be encoded to form the recovery information.

In the proposed scheme, each divided block can generate reserve room after LPC is executed. On the other hand, these divided blocks need auxiliary information to recover themselves, so the room is needed to store their auxiliary information. After calculating, we find that a few divided blocks reserve less room than they need to store their auxiliary information. So this kind of divided blocks are not suitable to be employed, and we refer to this kind of divided blocks as unused divided blocks and the remaining divided blocks as used divided blocks.

For unused divided blocks, we plan to keep their original pixel values unchanged and move them to the front area of the reserved image, so unused divided block information includes abscissa and ordinate of the unused divided blocks. Based on the previous assumptions, the divided blocks constitute an ($M/\tau \times N/\tau$) matrix, and each divided block in this matrix has an abscissa (ordinate) in the range of [$1, M/\tau$] ([$1, N/\tau$]). So the optimal length that can be used to represent the abscissa and ordinate binary value of an unused divided block is

$$d_0 = 2 \times \lceil \log_2 \max(M/\tau, N/\tau) \rceil. \qquad (5)$$

An example of encoding the auxiliary information of an unused divided block is shown in Fig. 5(a).

For most used divided blocks, their auxiliary information only contains region seeds; however, the auxiliary information of a few used divided blocks contains both region seeds and merged region information. According to our foregoing description, region seed is actually the $\lambda$ bits pixel of $I'_l$. Merged region information includes pixel value and position coordinate. The pixel value is actually the seed value of a merged region, so its length equal to $\lambda$. Position coordinate is the abscissa and ordinate of each pixel in the merged regions; its length can be given by

$$d_1 = 2 \times \lceil \log_2 \tau \rceil. \tag{6}$$

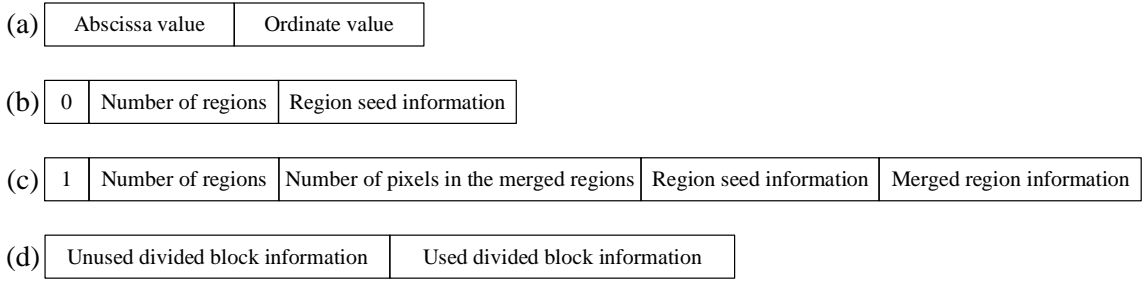

Fig. 5. Encoding rules of the auxiliary information. (a) Auxiliary information encoding rules for an unused divided block. (b) Auxiliary information encoding rules for a used divided block that contains no merged region. (c) Auxiliary information encoding rules for a used divided block that contains merged regions. (d) Encoding rules for the recovery information.

So the appropriate length for accommodating binary information of a merged pixel is

$$d_2 = d_1 + \lambda. \tag{7}$$

In Fig. 5(b) and Fig. 5(c), we design two data structures to encode the auxiliary information for used divided blocks with no merged region and used divided blocks with merged region respectively. In practice, we first generate auxiliary information of each individual used divided block respectively, and then connect them together. Please note that the length of the second part in Fig. 5(b) and Fig. 5(c) as well as the third part in Fig. 5(c) should be fixed values, which can make the encoded auxiliary information to be easily read. The way to determine these fixed lengths is to choose the maximum value that is needed. An example of the recovery information encoding rules is shown in Fig. 5(d).

Second, produce the reserved image ($I'_{re}$). In order to obtain a continuous available room, we do the following operations. First, in the raster order, we find the unused divided blocks (if any) and move them to the front area of $I'_{lc}$ one by one. Then, also in the raster order, the converted divided blocks are arranged into the remaining space of $I'_{lc}$. Finally, we store the recovery information generated above into the area immediately following the LSB bit plane of the unused divided blocks (if any). Now, the remaining available room in the low order bit plane of $I'_{lc}$ is a continuous room ($CR$) and we calculate each pixel value of $I'_{re}$ by

$$I'_{re}(a, o) = I'_h(a, o) + I'_{lc}(a, o). \tag{8}$$

After all the pixel values are calculated, $I'_{re}$ generates.

Finally, auxiliary information such as control parameters should be stored in the available room of $I'_{re}$. There are two types of control parameters: type-$\alpha$ and type-$\beta$. The type-$\alpha$ control parameter is the parameter related to image recovery, and it includes divided block size, the number of unused divided block, region seed binary value length as well as the storage location of the recovery information. The type-$\beta$ control parameter is a parameter related to data embedding and data extraction, and it only contains the starting coordinate of the available room. Assuming the number of the unused divided block is $\theta$, the binary length of type-$\alpha$ control parameter and type-$\beta$ control parameter can be given by Eq. (9) and Eq. (10), respectively.

$$d_3 = 3 \times \lceil \log_2 \max(\tau, \theta, \lambda) \rceil + 4 \times \lceil \log_2 \max(M, N) \rceil. \tag{9}$$

$$d_4 = 2 \times \lceil \log_2 \max(M, N) \rceil. \tag{10}$$

In the process of control parameters storing, we first set start mark $\acute{\sigma}$, which can only be recognized by the encryption key, at the beginning of $CR$, store the type-$\alpha$ control parameter into the $d_3$ bits next to $\acute{\sigma}$ in $CR$, and then set end mark $\acute{\sigma}$ immediately following the type-$\alpha$ control parameter. Finally, the type-$\beta$ control parameter is stored in the last $d_4$ bits of $CR$. Then the remaining room in $CR$, which we name as $CR'$, will be the available room that can be used to embed additional data. Suppose there are $\gamma$ used divided blocks in $I'_{re}$, the relationship between $\gamma$ and $\theta$ can be given as

$$\gamma = (M / \tau) \times (N / \tau) - \theta. \tag{11}$$

Then, the size of $CR'$ (represented by $\varphi$) can be calculated by

$$\varphi = \gamma \times \tau^2 \times (\lambda - 2) - \theta \times d_0 - \sum_{i=1}^{\gamma} (\delta_i \times \lambda + \varepsilon_i \times d_2) - d_3 - d_4, \tag{12}$$

where $\delta_i$ and $\varepsilon_i$ are the number of region seed and merged pixel in the $i$-th used division block.

In order to increase safety, the type-$\alpha$ parameter will be encrypted in the image encryption stage, and then it only can be accessed by people with the encryption key. The type-$\beta$ parameter will be encrypted in the data embedding stage, and then it only can be accessed by people with the data hiding key.

### C.  Image encryption

After the reserved image is generated, the content owner uses an encryption key to encrypt it, and the encryption method is stream cipher symmetric encryption. For an arbitrary pixel $I'_{re}(a, o)$ of $I'_{re}$, assuming its encryption version is $E(a, o)$, and $C(a, o)(k)$ represents the stream cipher, the encryption procedure contains two steps.

First, the content owner calculate each bit of $I'_{re}(a, o)$ with

$$I'_{re}(a,o)(k) = \left\lfloor \frac{I'_{re}(a,o)}{2^k} \right\rfloor \bmod 2, \quad k = 7, 6, 5, \cdots, 0. \quad (13)$$

Second, each bit of $I'_{re}(a, o)$ will be encrypted by using

$$E(a,o)(k) = I'_{re}(a,o)(k) \oplus C(a,o)(k), \quad k = 7, 6, 5, \cdots, 0. \quad (14)$$

With the exception of $CR'$ and type-$\beta$ parameter, all the content of $I'_{re}$ will be encrypted. Thus, anyone who does not has the encryption key cannot know the specific content of $I'_{re}$, she/he also cannot obtain the decrypted type-$\alpha$ parameter and therefore cannot recover the original image.

### D. Data embedding

The data embedding process contains four steps. First, after acquiring the encrypted image, data owner reads the type-$\beta$ control parameter, by which the data owner can know where she/he can start embedding additional data, from the last $d_d$ bits of the LSB bit plane of the encrypted image. Second, in order to increase security, the data owner encrypts additional data with a data hiding key. Third, the data embedding process is to replace the bits other than the last $d_d$ bits in $CR'$ with the encrypted additional data. When the embedding process is over, the data hider will set a mark $\vartheta$, which can only be recognized by the data hiding key, immediately following the encrypted additional data storage location to point out the end position of the embedding process. Finally, the data owner further encrypts the type-$\beta$ control parameter by using the data hiding key. So far, the marked encrypted image is generated. Anyone who does not has the data hiding key cannot know the specific content of the additional data, and she/he cannot even know where the embedded additional data starts and ends. The maximum embedding rate (represented by $\eta$) can be calculated by

$$\eta = \frac{\varphi}{M \times N}. \quad (15)$$

### E. Data extraction and image recovery

The receiver side operations are also shown in Fig. 2. As we can see, three situations are considered: a receiver only has the data hiding key, a receiver only has the encryption key, and a receiver has both the data hiding key and the encryption key.

**Case 1:** if a receiver only has the data hiding key, she/he can extract the additional data without error. First, she/he calculates the length of the type-$\beta$ control parameter by using Eq. (10), extracts the last $d_d$ bits of the LSB bit plane of the marked encrypted image and decrypts these extracted bits with the data hiding key, and then she/he can know where the encrypted additional data starts. Next, she/he starts to extract data from the starting coordinate until she/he encounters the end mark $\vartheta$. Finally, this receiver can accurately obtain the additional data by using the data hiding key to decrypt the extracted data.

**Case 2:** if a receiver only has the encryption key, she/he can obtain a high quality directly decrypted image. She/he first uses the encryption key to decrypt the marked encrypted image. In stream cipher symmetric encryption, decryption is the reverse process of encryption, so she/he can only decrypt the marked encrypted image space other than $CR'$ and the type-$\beta$ parameter, and this just guarantees that the previously embedded data and the type-$\beta$ parameter will not be changed. Next, she/he scans the decrypted marked image in the raster order until she/he encounters the start mark $\hat{o}$, then she/he extracts the following bits until she/he meets the end mark $\acute{o}$ to obtain the type-$\alpha$ control parameter. Third, with the storage location of the recovery information in the type-$\alpha$ control parameter, she/he can extract the recovery information from the decrypted marked image. With divided block size and the number of unused divided block in the type-$\alpha$ control parameter and Eq. (5), she/he can calculate the length of the unused divided block information and further divide the recovery information into unused divided block information (if any) and used divided block information. With divided block size and the number of unused divided block in the type-$\alpha$ control parameter, she/he can accurately extract unused divided block (if any) and used divided block from the decrypted marked image. Fourth, by using the unused divided block information (if any) to reposition each extracted unused divided block (if any) to its original position, the unused divided block can be recovered. Finally, she/he continues to use region division method to divide each converted divided block into regions. By using the region seed binary value length and the divided block size in the type-$\alpha$ control parameter, she/he extracts region seeds and merged region information (if any) from the used divided block information. Then, she/he replaces the covered value of each pixel with the extracted region seeds and recovers the merged region (if any). After the above operations are done, this receiver can obtain the high quality directly decrypted image.

**Case 3:** if a receiver has both the data hiding key and the encryption key, she/he can both extract additional data and recover the original image. First, with the data hiding key, she/he accurately extracts the additional data by using the way we described in case 1. Second, she/he decrypts the extracted encrypted image with the encryption key. Finally, without the embedded additional data, she/he can recover the entire image space, including the reserved available room in the way we described in case2, and then this receiver obtains the original image losslessly.

## IV. EXPERIMENT AND ANALYSIS

In this section, we first analyze the auxiliary information size, encryption security and time complexity of the proposed scheme. Next, we give the specific execution results of the proposed scheme. Finally, we compare the proposed scheme with the state-of-the-art algorithms [18]-[24]. The experiments below are based on publicly available standard grayscale images datasets [33].

### A. Auxiliary information size analysis

Tables II-IV show the size of the auxiliary information and the size of the total available room produced during the execution of the proposed scheme when $\tau \times \tau$ is equal to 32×32, 16×16, and 8×8, respectively. The data in Tables II-IV is obtained by testing the 9 images of 512×512×8 size with different texture complexity and then calculating the test results with Eqs. (5)-(7) and Eqs. (9)-(10). By analyzing the data in Tables II-IV, three points are reflected.

First of all, the size of the total available room reserved is much larger than the size of the total auxiliary information generated. This is due to the setting of unused divided blocks, which we described earlier.

Second, a large $\tau\times\tau$ value will lead to a high embedding capacity. This is because a larger divided block contains more pixels, which will help reduce the destruction of the redundancy space we made by image pre-processing. It is this feature that leads to the reduction of the size of the total auxiliary information. Therefore, a high embedding capacity can be obtained.

Third, images with simpler texture will have a higher embedding capacity than that of the complex texture images. For the different images, when the value of $\tau\times\tau$ is the same, images with simpler texture will generate a smaller size of auxiliary information. This can help increase the size of the reserved room and further generate a high embedding capacity.

### B. Encryption security analysis

The security of the encryption used in the proposed scheme will be analyzed in the follows. In order to do this, a comparison is shown in Fig. 6, and the comparison items of horizontal-direction adjacent pixel differences and vertical-direction adjacent pixel differences are respectively calculated by

$$\psi_h(i, j) = I''(i, j) - I''(i, j+1), 1 \le i \le M, 1 \le j \le N-1, \text{(16)}$$

and

$$\psi_v(i, j) = I''(i, j) - I''(i+1, j), 1 \le i \le M-1, 1 \le j \le N, \text{(17)}$$

where $\psi_h(i, j)$ is the horizontal-direction difference between two adjacent pixels, $\psi_v(i, j)$ is the vertical-direction difference between two adjacent pixels, and $I''(i, j)$ represents the image pixel to be calculated. The comparison in Fig. 6 reflects that the algorithm is safe, and the reasons are as follows.

TABLE II The Size of Auxiliary Information and Total Available Room Reserved When $\tau\times\tau$ is Equal to 32×32.

| Image | Sub-size of each auxiliary information(bit) | | | | Total auxiliary information | | Total available room reserved | |
|---|---|---|---|---|---|---|---|---|
| | Unused divided block information | Used divided block information | Type-α control parameter | Type-β control parameter | Size(bit) | Payload(bbp) | Size(bit) | Payload(bbp) |
| Airplane | 0 | 28098 | 51 | 18 | 28167 | 0.1074 | 262144 | 1.0000 |
| Peppers | 0 | 28624 | 51 | 18 | 28693 | 0.1095 | 262144 | 1.0000 |
| Lena | 0 | 28444 | 51 | 18 | 28513 | 0.1088 | 262144 | 1.0000 |
| Crowd | 8 | 32335 | 51 | 18 | 32412 | 0.1236 | 261120 | 0.9961 |
| Boat | 24 | 39399 | 51 | 18 | 39492 | 0.1507 | 259072 | 0.9883 |
| Lake | 8 | 51197 | 51 | 18 | 51274 | 0.1956 | 261120 | 0.9961 |
| Man | 56 | 51708 | 51 | 18 | 51833 | 0.1977 | 254976 | 0.9727 |
| Barbara | 256 | 39059 | 51 | 18 | 39384 | 0.1502 | 229376 | 0.8750 |
| Baboon | 456 | 67540 | 54 | 18 | 68068 | 0.2597 | 203776 | 0.7773 |

TABLE III The Size of Auxiliary Information and Total Available Room Reserved When $\tau\times\tau$ is Equal to 16×16.

| Image | Sub-size of each auxiliary information(bit) | | | | Total auxiliary information | | Total available room reserved | |
|---|---|---|---|---|---|---|---|---|
| | Unused divided block information | Used divided Block information | Type-α control parameter | Type-β control parameter | Size(bit) | Payload(bbp) | Size(bit) | Payload(bbp) |
| Airplane | 10 | 36211 | 48 | 18 | 36287 | 0.1384 | 261888 | 0.9990 |
| Peppers | 0 | 39878 | 48 | 18 | 39944 | 0.1524 | 262144 | 1.0000 |
| Lena | 10 | 37352 | 48 | 18 | 37428 | 0.1428 | 261888 | 0.9990 |
| Crowd | 0 | 42691 | 48 | 18 | 42757 | 0.1631 | 262144 | 1.0000 |
| Boat | 10 | 46714 | 48 | 18 | 46790 | 0.1785 | 261888 | 0.9990 |
| Lake | 40 | 55855 | 48 | 18 | 55961 | 0.2135 | 261120 | 0.9961 |
| Man | 90 | 64172 | 48 | 18 | 64328 | 0.2454 | 259840 | 0.9912 |
| Barbara | 820 | 53292 | 57 | 18 | 54187 | 0.2067 | 241152 | 0.9199 |
| Baboon | 1090 | 93993 | 57 | 18 | 95158 | 0.3630 | 234240 | 0.8936 |

TABLE IV The Size of Auxiliary Information and Total Available Room Reserved When $\tau\times\tau$ is Equal to 8×8.

| Image | Sub-size of each auxiliary information(bit) | | | | Total auxiliary information | | Total available room reserved | |
|---|---|---|---|---|---|---|---|---|
| | Unused divided block information | Used divided Block information | Type-α control parameter | Type-β control parameter | Size(bit) | Payload(bbp) | Size(bit) | Payload(bbp) |
| Airplane | 36 | 69933 | 45 | 18 | 70032 | 0.2672 | 261952 | 0.9993 |
| Peppers | 60 | 75318 | 45 | 18 | 75441 | 0.2878 | 261824 | 0.9988 |
| Lena | 60 | 71856 | 45 | 18 | 71979 | 0.2746 | 261824 | 0.9988 |
| Crowd | 60 | 79095 | 45 | 18 | 79218 | 0.3022 | 261824 | 0.9988 |
| Boat | 96 | 82824 | 45 | 18 | 82983 | 0.3166 | 261632 | 0.9980 |
| Lake | 168 | 92612 | 48 | 18 | 92846 | 0.3542 | 261248 | 0.9966 |
| Man | 516 | 99828 | 54 | 18 | 100416 | 0.3831 | 259392 | 0.9895 |
| Barbara | 3312 | 91193 | 63 | 18 | 94586 | 0.3608 | 244480 | 0.9326 |
| Baboon | 5184 | 128490 | 63 | 18 | 133755 | 0.5102 | 234496 | 0.8945 |

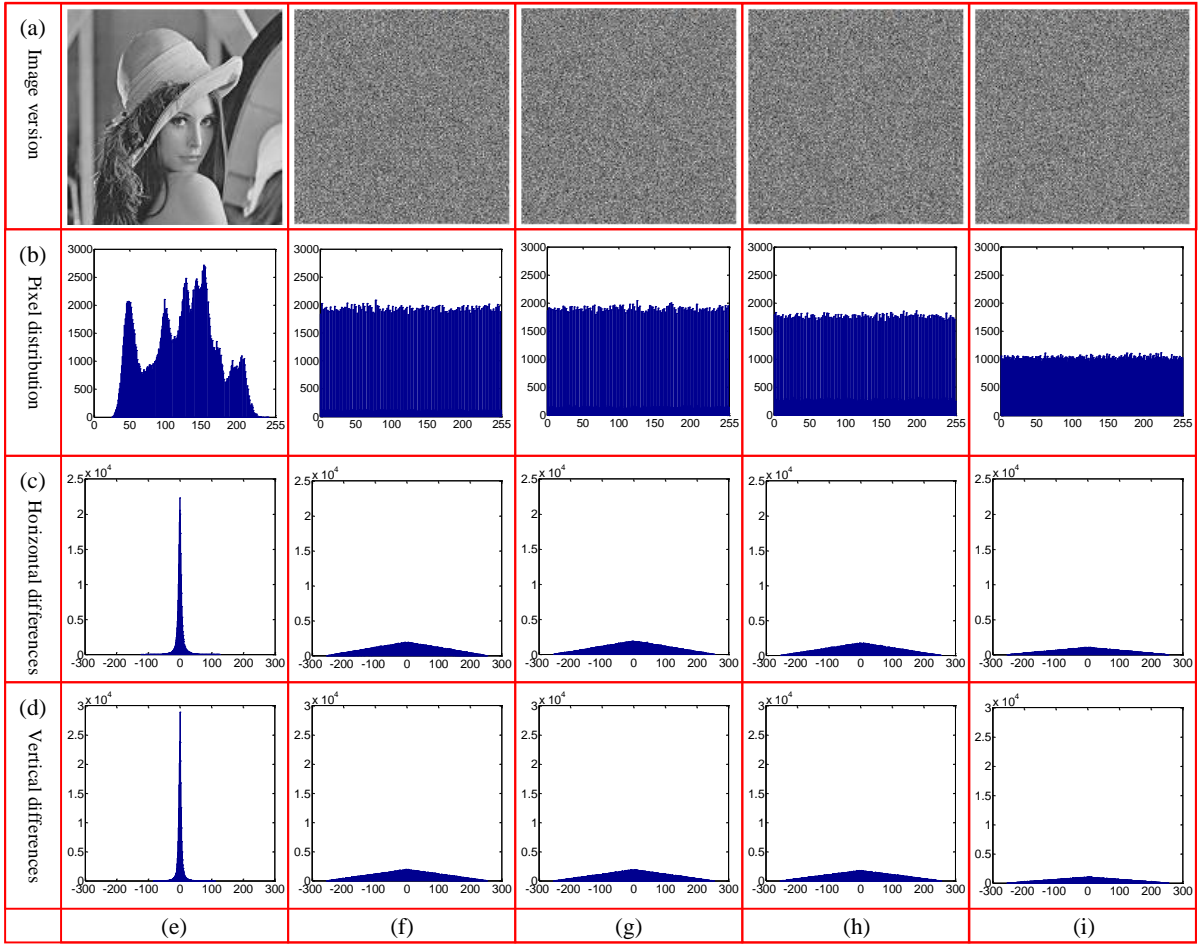

Fig. 6. Comparison of the statistical characteristics of the original image, the encrypted image produced by the proposed scheme, and the encrypted image produced by the standard stream cipher encryption method. (a, e) is the original image. (a, f), (a, g), and (a, h) are the encrypted images produced by the proposed scheme when $\tau \times \tau$ is set to 32×32, 16×16, and 8×8, respectively. (a, i) is the encrypted image produced by the standard stream cipher encryption method. Row (b) shows the pixel values distribution of row (a), row (c) shows the horizontal-direction adjacent pixel differences distribution of row (a), and row (d) is the vertical-direction adjacent pixel differences distribution of row (a).

First, pixel values distribution in the encrypted image produced by the proposed scheme is uniform. As is shown, pixel values distribution in (b, e) of Fig. 6 is uneven; however, the situation is the opposite in (b, f), (b, g), and (b, h) of Fig. 6. For the three different values of $\tau \times \tau$, the pixel values in the produced encrypted image are all evenly distributed, and the smaller the value of $\tau \times \tau$, the closer the distribution of pixel values is to that of (b, i) of Fig. 6, which is the encrypted image produced by the standard stream cipher encryption method.

Second, in the encrypted image produced by the proposed scheme, adjacent pixel differences distributions in both the horizontal-direction and the vertical-direction are very different from that of the original image, but very similar to that of the encrypted image produced by the standard stream cipher encryption method. As can be seen, both (c, e) and (d, e) in Fig. 6 are centrally distributed and have large peak values. However, adjacent pixel differences in (c, f-h) and (d, f-h) of Fig. 6 are distributed across the entire data range and only have small peak values, and those distribution characteristics are very similar to (c, i) and (d, i) of Fig. 6.

### C. Time complexity analysis and comparison

This section analyzes the time complexity of the proposed scheme from the angle of prior analysis, and then compares it with that of the other similar algorithms. In the analysis and comparison, we assume the size of the input image is $M \times N$, algorithm $\Lambda$ divides the input image into $\chi_\Lambda$ divided blocks (if algorithm $\Lambda$ contains division operation) and embeds $\Delta_\Lambda$ bits additional data. Data extraction is the reverse process of data embedding, and they have the same time complexity. Image recovery is the reverse process of reserving available room and image encryption, and they also have the same time complexity. Therefore, for direct and concise comparison, the following analysis and comparison only consider the time complexity of the three stages of reserving available room, image encryption and data embedding.

**Time complexity analysis**. The reserve available room stage in the proposed scheme contains LPC and reserved image generation. Image pre-processing in LPC processes the pixels of the input image one by one, so it has a time complexity of $O(M \times N)$. The following phases of LPC process all the divided blocks in turn, so they have a time complexity of $O(\chi_{proposed})$. Reserved image generation also processes the divided blocks one by one, so it also has a time complexity of $O(\chi_{proposed})$. Image encryption of the proposed scheme encrypts each pixel of the reserved image, so it has a time complexity of $O(M \times N)$. Data embedding of the proposed scheme only operates the pixels used to store additional data, so it has a time complexity of $O(\Delta_{proposed})$. In the same way, we also analyze the time complexity of other similar algorithms. Time complexity analysis results are shown in Table V.

**Time complexity comparison**. If $\Delta_{Ma[18]}$ and $\Delta_{proposed}$ are the same size, the three stages time complexities of the proposed scheme in Table V will obviously smaller than that of Ma[18]. For the remaining comparison, we set the divided bock sizes in the proposed scheme and Li[23] are respectively $32 \times 32$ and $5 \times 5$, which will allow them to achieve their good performance, and then $\chi_{Li[23]}$ equals to $M \times N/25$ and $\chi_{proposed}$ equals to $M \times N/1024$. If W. Zhang[19], Li[23] and the proposed scheme embed the same size additional data, we can find that the proposed scheme still has an advantage in terms of time complexity compared to W. Zhang[19] and Li[23] by substituting $\chi_{Li[23]}$ and $\chi_{proposed}$ into Table V. On the other hand, although the total time complexities of the three stages of the algorithms in Table V are all in the same order of magnitude, the proposed scheme still has a lower time complexity when the scale of the input image is constant.

### D. Individual execution results

In this section, we first use a standard test image, Lena, to perform the whole process of our proposed scheme. Next, we execute a depth test by using more images and different block sizes.

#### 1) The whole process of the proposed scheme

The results of the key phases of the proposed scheme are shown in Fig. 7, in which the divided block size ($\tau \times \tau$) is set to $32 \times 32$. In Fig. 7(e), a payload of 0.89 bpp additional data was embedded, and its PSNR value compared with Fig. 7(a) is 51.65 dB. In Fig. 7(f), the recovered image has a PSNR value of *Inf* dB. According to the results in Fig. 7, the advantage of our proposed scheme such as high PSNR value of directly decrypted image can be verified.

#### 2) Depth test of the proposed scheme

In Table VI, we use nine different styles of images to test the proposed scheme. The impact of the divided block size ($\tau \times \tau$) on the performance of the proposed scheme is also taken into consideration, and we use three different divided block sizes, $32 \times 32$, $16 \times 16$ and $8 \times 8$ to test this. As we can see, for the same image, $\eta$ increases as $\tau \times \tau$ becomes larger, while PSNR becomes lower. The reason is a larger block will keep the redundancy space from being decomposed. On the other hand, if $\tau \times \tau$ keeps unchanged, the value of $\eta$ will increase as the image texture becomes simpler; however, the value of PSNR will decrease. This is because simple texture images need less recovery information to recover themselves, which will cause the reserved room size to become larger. Thus, $\eta$ will increase, and then a higher embedding capacity corresponds to the lower PSNR value. Actually, we have inferred the above results when we analyzed Tables II-IV, and the forecasts are verified here. It should be noted that, for all the experiments in Table VI, the data extraction correct rate is 100% and the PSNR of the recovered image is infinity.

TABLE V RESULTS OF TIME COMPLEXITY ANALYSIS.

| Algorithm | Time complexity of | | |
| --- | --- | --- | --- |
| | Reserve available room | Image encryption | Data embedding |
| Ma[18] | $O(M \times N) + O(\chi_{Ma[18]}) + O(2 \times M \times N \times (\chi_{Ma[18]}-1)/\chi_{Ma[18]})$ | $O(M \times N)$ | $O(\Delta_{Ma[18]})$ |
| W.zhang[19] | $O(0.8 \times M \times N) + O(M \times N)$ | $O(0.8 \times M \times N)$ | $O(0.2 \times M \times N)$ |
| Li[23] | $O(\chi_{Li[23]}) + O(5 \times \chi_{Li[23]} \times 5)$ | $O(\chi_{Li[23]}) + O(M \times N)$ | $O(5 \times \chi_{Li[23]} \times 3)$ |
| Proposed | $O(M \times N) + O(\chi_{proposed}) + O(\chi_{proposed})$ | $O(M \times N)$ | $O(\Delta_{proposed})$ |

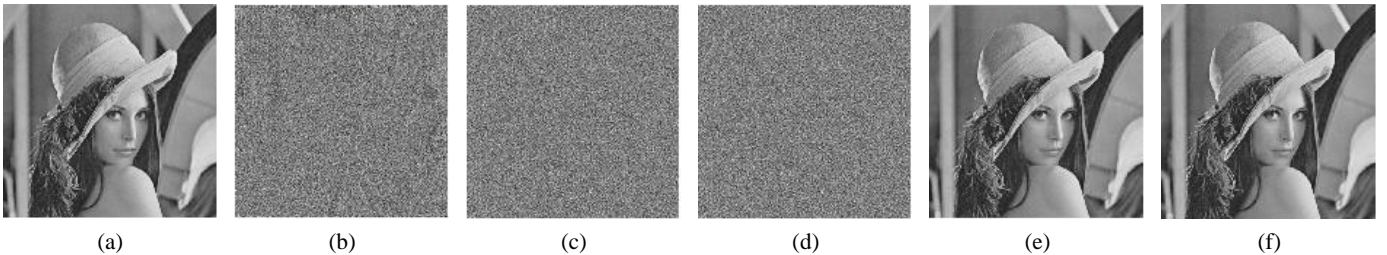

(a)        (b)        (c)        (d)        (e)        (f)

Fig. 7. Key phase results of the proposed scheme. (a) The original image. (b) The reserved image. (c) The encrypted image. (d) The marked encrypted image. (e) The directly decrypted image. (f) The recovered image.

TABLE VI TESTING RESULTS FOR DIFFERENT IMAGES UNDER DIFFERENT $\tau \times \tau$.

| $\tau \times \tau$ | $\eta$(bpp) | | | PSNR(dB) | | |
|---|---|---|---|---|---|---|
| | 32×32 | 16×16 | 8×8 | 32×32 | 16×16 | 8×8 |
| Airplane | 0.89 | 0.86 | 0.73 | 51.66 | 51.80 | 52.51 |
| Peppers | 0.89 | 0.85 | 0.71 | 51.66 | 51.87 | 52.62 |
| Lena | 0.89 | 0.86 | 0.72 | 51.65 | 51.83 | 52.54 |
| Crowd | 0.87 | 0.84 | 0.70 | 51.75 | 51.93 | 52.70 |
| Boat | 0.83 | 0.82 | 0.68 | 51.94 | 52.00 | 52.80 |
| Lake | 0.79 | 0.78 | 0.64 | 52.13 | 52.19 | 53.07 |
| Man | 0.77 | 0.74 | 0.61 | 52.29 | 52.44 | 53.33 |
| Barbara | 0.72 | 0.71 | 0.57 | 52.57 | 52.62 | 53.58 |
| Baboon | 0.51 | 0.53 | 0.38 | 54.07 | 53.90 | 55.33 |

### E. Comparison with other algorithms

In this section, we compare the proposed scheme with the state-of-the-art algorithms, in which the comparison items include embedding rate and PSNR value of the directly decrypted image. In the following comparisons, we choose the divided block size as 32×32 to make the proposed scheme have a better performance. In order to be fair, we also make the compared algorithms in their best performance.

#### 1) Comparison of testing individual image

Fig. 8 plots the comparison results, from which we can see that our proposed scheme outperforms all the compared algorithms for all the four test images. Although some compared algorithms can reach an equivalent embedding rate as our proposed scheme, our proposed scheme has a higher PSNR value. Thus, compared with the other algorithms, the proposed scheme always keeps superiority in performance.

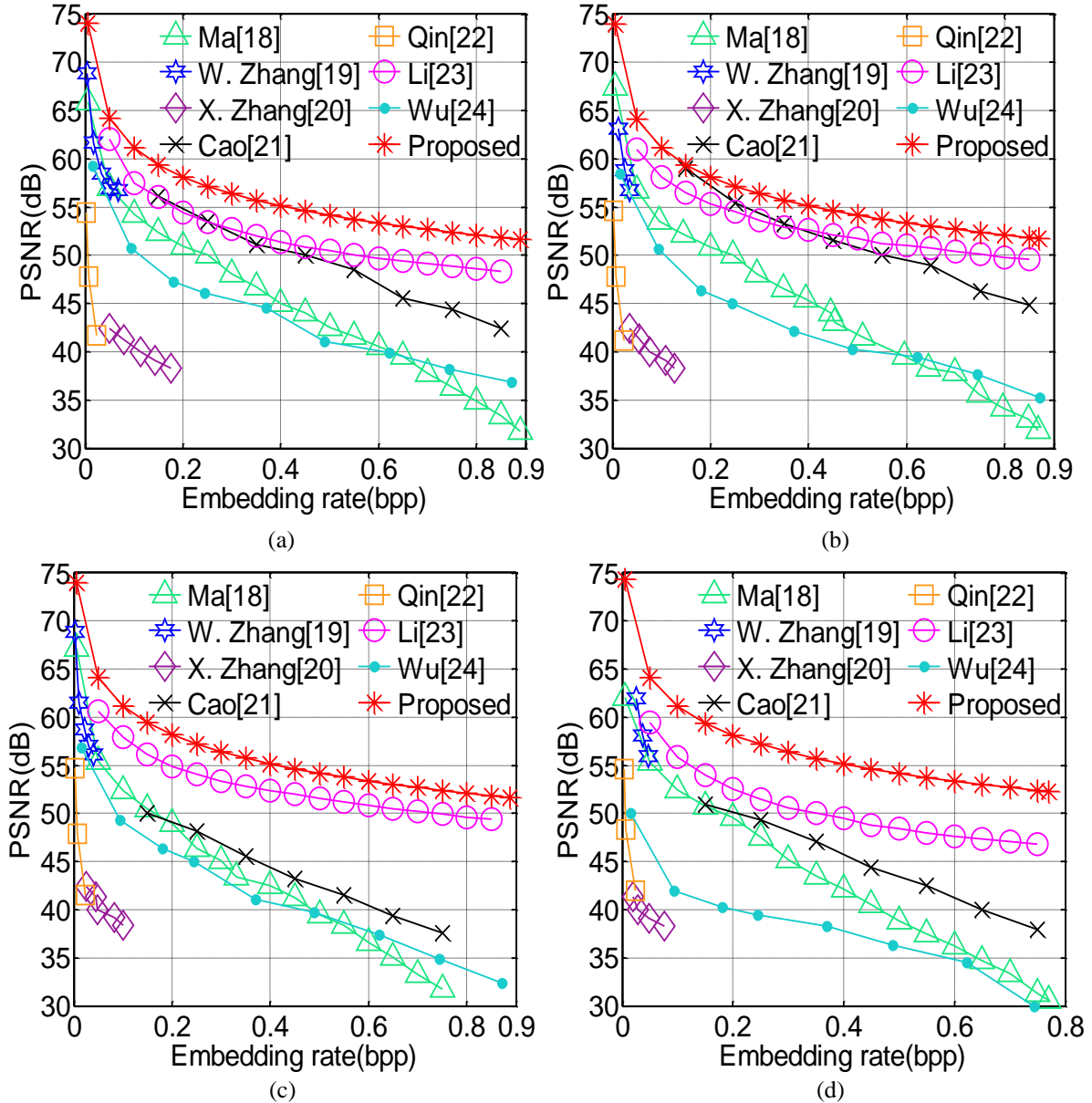

Fig. 8. PSNR comparison of the directly decrypted image. (a) Airplane. (b) Crowd. (c) Lena. (d) Man.

*2) Comparison of average performance*

The comparison results of average performance are shown in Fig. 9, in which the results are obtained by averaging the test results of 800 standard images of [41]. In Fig. 9, the proposed scheme can reach a high embedding rate of near to 1 bpp, while most compared algorithms cannot. Moreover, compared with Ma [18], Cao [21], Li [23], and Wu [24], the proposed scheme has a higher PSNR value when the embedding rate is the same. So, in general, the average performance of the proposed scheme still has advantages.

Actually, the average performance of the proposed scheme in Fig. 9 is universal, and the reasons contain two points. First, the test images used in the comparison experiment are diverse. The tested image styles include the biomedical image, astronomical image, marble image, illusory image, contour image and miscellaneous image, and the sizes of the tested images contain 512×512×8, 256×256×8, and 128×128×8. These diversities will ensure the universality of the test results. Second, the principle of the proposed scheme is to divide the image into regions based on the spatial correlation of natural images, and then generate available room from the divided regions to store additional data. The universal existence of spatial correlation also provides a guarantee for the universality of the performance of the proposed scheme.

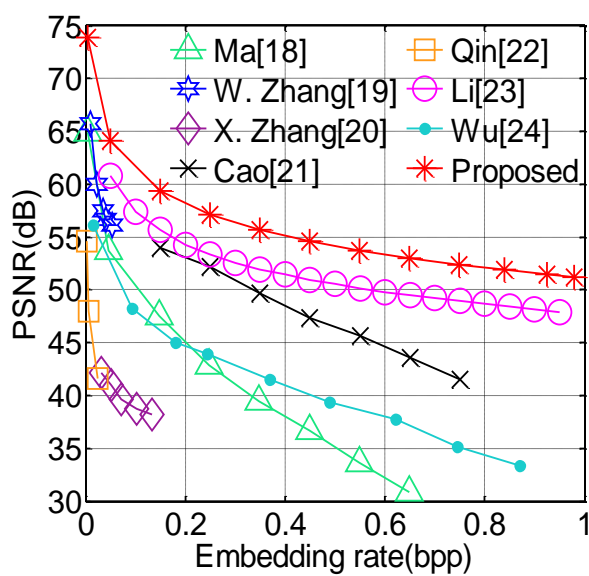

Fig. 9. Comparison of average performance.

## V. Conclusion

Recently, many researchers have proposed algorithms for RDHEI. However, these algorithms only have a low embedding capacity, so this paper proposes a novel RDHEI scheme by lossless pixel conversion (LPC). In general, the novelty of the proposed scheme is reflected in three aspects: inspired by the planar map coloring question, divide the original image into irregular regions, and dynamic image division process. In LPC, we extract the redundant space in the original image according to the spatial correlation in natural images, and obtain available room for accommodating additional data by reversible converting the pixel value in the redundant space to another form. In the proposed scheme, we first reserve room in the original image by LPC, and then encrypt the reserved image with a standard stream cipher encryption method. Next, we embed additional data into the reserved room of the encrypted image. Finally, data extraction and image recovery are separable operations on the receiver side. Experimental results show that the proposed scheme has a good performance. In the future, we will further improve the embedding capacity of the proposed scheme.